\documentclass[prd,superscriptaddress,showpacs,amssymb,amsmath,amsfonts,aps,altaffilletter,nofootinbib,letterpaper,twocolumn]{revtex4}

\usepackage{color}
\usepackage{graphicx}
\usepackage{fancyhdr}
\usepackage{float}
\usepackage[raggedright]{subfigure}
\usepackage{acronym}
\usepackage{multirow}
\usepackage{array}
\usepackage{hyperref}

\makeatletter
\makeatletter
\def\@fnsymbol#1{\ifcase#1\or * \or  $+$ \or  \$ \or \#  \or \dag \or \ddag \or
$\mathsection$ \or $ \mathsubsection$ \or $\|$  \or \textordfeminine \or \textbullet   
\or ** \or $++$ \or  \$\$ \or \#\#  \or \dag\dag \or \ddag\ddag \or
$\mathsection\mathsection$ \or $ \mathsubsection\mathsubsection$ \or $\|\|$  \or 
\textordfeminine\textordfeminine \or \textbullet \textbullet \or *** \or $+++$ 
\or  \$\$\$ \or \#\#  \or \dag\dag \or \ddag\ddag \or
$\mathsection \mathsection\mathsection$ \or $ \mathsubsection 
\mathsubsection\mathsubsection$ \or $\|\|\|$  \or 
\textordfeminine\textordfeminine\textordfeminine \or 
\textbullet\textbullet\textbullet \or \else \@ctrerr\fi}
\makeatother

\newcommand\fake[1]{\textcolor{red}{#1}}

\def\thercsid{\relax}
\def\rcsid#1{\def\next##1#1{\def\thercsid{##1}}\next}

\rcsid$Id: s6-highmass.tex,v 1.100 2013/01/09 16:11:17 thomas.dent Exp $

\renewcommand{\today}{\number\day\space\ifcase\month\or
  January\or February\or March\or April\or May\or June\or
  July\or August\or September\or October\or November\or December\fi
  \space\number\year}

\newcommand\perMpcyr{\ensuremath{\mathrm{Mpc}^{-3}\mathrm{yr}^{-1}}}
\def\Msun{\ensuremath{\mathrm{M_{\odot}}}}

\def\ssixstart{July 7, 2009}
\def\ssixend{October 20, 2010}
\def\twentytwentyhorizon{300\,Mpc}
\def\twentythreetwentythreelimit{$3.3\times 10^{-7}$ mergers \perMpcyr}
\def\twentythreefourteenlimit{$5.9\times 10^{-7}$ mergers \perMpcyr}

\def\loudestGPS{939789782}
\def\secondloudestGPS{963363786}
\def\loudestFAR{$0.41\,\mathrm{yr}^{-1}$}
\def\secondFAR{$1.0\,\mathrm{yr}^{-1}$}
\def\loudestRhoc{\ensuremath{11.98}}
\def\secondRhoc{\ensuremath{10.48}}

\def\totalLoudTime{\ensuremath{0.53}\,yr}
\def\totalpostVeto{\ensuremath{0.47}\,yr}
\def\HLlt{\ensuremath{0.17}\,yr}
\def\HVlt{\ensuremath{0.10}\,yr}
\def\LVlt{\ensuremath{0.07}\,yr}
\def\HLVlt{\ensuremath{0.09}\,yr}
\def\totalTime{\ensuremath{0.42}\,yr}

\begin{document}

\title{Search for Gravitational Waves from 
Binary Black Hole 
Inspiral, Merger and Ringdown 
in LIGO-Virgo Data from 2009--2010
}

%
%
%

\affiliation{LIGO - California Institute of Technology, Pasadena, CA  91125, USA}
\affiliation{California State University Fullerton, Fullerton CA 92831 USA}
\affiliation{SUPA, University of Glasgow, Glasgow, G12 8QQ, United Kingdom}
\affiliation{Laboratoire d'Annecy-le-Vieux de Physique des Particules (LAPP), Universit\'e de Savoie, CNRS/IN2P3, F-74941 Annecy-Le-Vieux, France}
\affiliation{INFN, Sezione di Napoli $^a$; Universit\`a di Napoli 'Federico II'$^b$, Complesso Universitario di Monte S.Angelo, I-80126 Napoli; Universit\`a di Salerno, Fisciano, I-84084 Salerno$^c$, Italy}
\affiliation{LIGO - Livingston Observatory, Livingston, LA  70754, USA}
\affiliation{Cardiff University, Cardiff, CF24 3AA, United Kingdom}
\affiliation{University of Sannio at Benevento, I-82100 Benevento, Italy and INFN (Sezione di Napoli), Italy}
\affiliation{Albert-Einstein-Institut, Max-Planck-Institut f\"ur Gravitationsphysik, D-30167 Hannover, Germany}
\affiliation{Leibniz Universit\"at Hannover, D-30167 Hannover, Germany}
\affiliation{Nikhef, Science Park, Amsterdam, The Netherlands$^a$; VU University Amsterdam, De Boelelaan 1081, 1081 HV Amsterdam, The Netherlands$^b$}
\affiliation{National Astronomical Observatory of Japan, Tokyo  181-8588, Japan}
\affiliation{University of Wisconsin--Milwaukee, Milwaukee, WI  53201, USA}
\affiliation{INFN, Sezione di Pisa$^a$; Universit\`a di Pisa$^b$; I-56127 Pisa; Universit\`a di Siena, I-53100 Siena$^c$, Italy}
\affiliation{University of Florida, Gainesville, FL  32611, USA}
\affiliation{INFN, Sezione di Roma$^a$; Universit\`a 'La Sapienza'$^b$, I-00185 Roma, Italy}
\affiliation{LIGO - Hanford Observatory, Richland, WA  99352, USA}
\affiliation{University of Birmingham, Birmingham, B15 2TT, United Kingdom}
\affiliation{Albert-Einstein-Institut, Max-Planck-Institut f\"ur Gravitationsphysik, D-14476 Golm, Germany}
\affiliation{Montana State University, Bozeman, MT 59717, USA}
\affiliation{European Gravitational Observatory (EGO), I-56021 Cascina (PI), Italy}
\affiliation{Syracuse University, Syracuse, NY  13244, USA}
\affiliation{LIGO - Massachusetts Institute of Technology, Cambridge, MA 02139, USA}
\affiliation{APC, AstroParticule et Cosmologie, Universit\'e Paris Diderot, CNRS/IN2P3, CEA/Irfu, Observatoire de Paris, Sorbonne Paris Cit\'e, 10, rue Alice Domon et L\'eonie Duquet, 75205 Paris Cedex 13, France}
\affiliation{Columbia University, New York, NY  10027, USA}
\affiliation{Stanford University, Stanford, CA  94305, USA}
\affiliation{IM-PAN 00-956 Warsaw$^a$; Astronomical Observatory Warsaw University 00-478 Warsaw$^b$; CAMK-PAN 00-716 Warsaw$^c$; Bia{\l}ystok University 15-424 Bia{\l}ystok$^d$; NCBJ 05-400 \'Swierk-Otwock$^e$; Institute of Astronomy 65-265 Zielona G\'ora$^f$,  Poland}
\affiliation{The University of Texas at Brownsville, Brownsville, TX 78520, USA}
\affiliation{San Jose State University, San Jose, CA 95192, USA}
\affiliation{Moscow State University, Moscow, 119992, Russia}
\affiliation{LAL, Universit\'e Paris-Sud, IN2P3/CNRS, F-91898 Orsay$^a$; ESPCI, CNRS,  F-75005 Paris$^b$, France}
\affiliation{NASA/Goddard Space Flight Center, Greenbelt, MD  20771, USA}
\affiliation{University of Western Australia, Crawley, WA 6009, Australia}
\affiliation{The Pennsylvania State University, University Park, PA  16802, USA}
\affiliation{Universit\'e Nice-Sophia-Antipolis, CNRS, Observatoire de la C\^ote d'Azur, F-06304 Nice$^a$; Institut de Physique de Rennes, CNRS, Universit\'e de Rennes 1, 35042 Rennes$^b$, France}
\affiliation{Laboratoire des Mat\'eriaux Avanc\'es (LMA), IN2P3/CNRS, F-69622 Villeurbanne, Lyon, France}
\affiliation{Washington State University, Pullman, WA 99164, USA}
\affiliation{INFN, Sezione di Perugia$^a$; Universit\`a di Perugia$^b$, I-06123 Perugia; Universit\`a di Camerino, Dipartimento di Fisica$^c$, I-62032 Camerino, Italy}
\affiliation{INFN, Sezione di Firenze, I-50019 Sesto Fiorentino$^a$; Universit\`a degli Studi di Urbino 'Carlo Bo', I-61029 Urbino$^b$, Italy}
\affiliation{University of Oregon, Eugene, OR  97403, USA}
\affiliation{Laboratoire Kastler Brossel, ENS, CNRS, UPMC, Universit\'e Pierre et Marie Curie, 4 Place Jussieu, F-75005 Paris, France}
\affiliation{University of Maryland, College Park, MD 20742 USA}
\affiliation{Universitat de les Illes Balears, E-07122 Palma de Mallorca, Spain}
\affiliation{University of Massachusetts - Amherst, Amherst, MA 01003, USA}
\affiliation{Canadian Institute for Theoretical Astrophysics, University of Toronto, Toronto, Ontario, M5S 3H8, Canada}
\affiliation{Tsinghua University, Beijing 100084 China}
\affiliation{University of Michigan, Ann Arbor, MI  48109, USA}
\affiliation{Louisiana State University, Baton Rouge, LA  70803, USA}
\affiliation{The University of Mississippi, University, MS 38677, USA}
\affiliation{Charles Sturt University, Wagga Wagga, NSW 2678, Australia}
\affiliation{Caltech-CaRT, Pasadena, CA  91125, USA}
\affiliation{INFN, Sezione di Genova;  I-16146  Genova, Italy}
\affiliation{Pusan National University, Busan 609-735, Korea}
\affiliation{Australian National University, Canberra, ACT 0200, Australia}
\affiliation{Carleton College, Northfield, MN  55057, USA}
\affiliation{The University of Melbourne, Parkville, VIC 3010, Australia}
\affiliation{INFN, Sezione di Roma Tor Vergata$^a$; Universit\`a di Roma Tor Vergata, I-00133 Roma$^b$; Universit\`a dell'Aquila, I-67100 L'Aquila$^c$, Italy}
\affiliation{University of Salerno, I-84084 Fisciano (Salerno), Italy}
\affiliation{Instituto Nacional de Pesquisas Espaciais,  12227-010 - S\~{a}o Jos\'{e} dos Campos, SP, Brazil}
\affiliation{The University of Sheffield, Sheffield S10 2TN, United Kingdom}
\affiliation{Wigner RCP, RMKI, H-1121 Budapest, Konkoly Thege Mikl\'os \'ut 29-33, Hungary}
\affiliation{Inter-University Centre for Astronomy and Astrophysics, Pune - 411007, India}
\affiliation{University of Minnesota, Minneapolis, MN 55455, USA}
\affiliation{INFN, Gruppo Collegato di Trento$^a$ and Universit\`a di Trento$^b$,  I-38050 Povo, Trento, Italy;   INFN, Sezione di Padova$^c$ and Universit\`a di Padova$^d$, I-35131 Padova, Italy}
\affiliation{California Institute of Technology, Pasadena, CA  91125, USA}
\affiliation{Northwestern University, Evanston, IL  60208, USA}
\affiliation{Rochester Institute of Technology, Rochester, NY  14623, USA}
\affiliation{E\"otv\"os Lor\'and University, Budapest, 1117 Hungary}
\affiliation{University of Cambridge, Cambridge, CB2 1TN, United Kingdom}
\affiliation{University of Szeged, 6720 Szeged, D\'om t\'er 9, Hungary}
\affiliation{Rutherford Appleton Laboratory, HSIC, Chilton, Didcot, Oxon OX11 0QX United Kingdom}
\affiliation{Embry-Riddle Aeronautical University, Prescott, AZ  86301 USA}
\affiliation{Perimeter Institute for Theoretical Physics, Ontario, N2L 2Y5, Canada}
\affiliation{American University, Washington, DC 20016, USA}
\affiliation{University of New Hampshire, Durham, NH 03824, USA}
\affiliation{University of Southampton, Southampton, SO17 1BJ, United Kingdom}
\affiliation{Korea Institute of Science and Technology Information, Daejeon 305-806, Korea}
\affiliation{Hobart and William Smith Colleges, Geneva, NY  14456, USA}
\affiliation{Institute of Applied Physics, Nizhny Novgorod, 603950, Russia}
\affiliation{Lund Observatory, Box 43, SE-221 00, Lund, Sweden}
\affiliation{Hanyang University, Seoul 133-791, Korea}
\affiliation{Seoul National University, Seoul 151-742, Korea}
\affiliation{University of Strathclyde, Glasgow, G1 1XQ, United Kingdom}
\affiliation{The University of Texas at Austin, Austin, TX 78712, USA}
\affiliation{Southern University and A\&M College, Baton Rouge, LA  70813, USA}
\affiliation{University of Rochester, Rochester, NY  14627, USA}
\affiliation{University of Adelaide, Adelaide, SA 5005, Australia}
\affiliation{National Institute for Mathematical Sciences, Daejeon 305-390, Korea}
\affiliation{Louisiana Tech University, Ruston, LA  71272, USA}
\affiliation{McNeese State University, Lake Charles, LA 70609 USA}
\affiliation{Andrews University, Berrien Springs, MI 49104 USA}
\affiliation{Trinity University, San Antonio, TX  78212, USA}
\affiliation{University of Washington, Seattle, WA, 98195-4290, USA}
\affiliation{Southeastern Louisiana University, Hammond, LA  70402, USA}

\author{J.~Aasi$^\text{1}$}\noaffiliation
\author{J.~Abadie$^\text{1}$}\noaffiliation
\author{B.~P.~Abbott$^\text{1}$}\noaffiliation
\author{R.~Abbott$^\text{1}$}\noaffiliation
\author{T.~D.~Abbott$^\text{2}$}\noaffiliation
\author{M.~Abernathy$^\text{3}$}\noaffiliation
\author{T.~Accadia$^\text{4}$}\noaffiliation
\author{F.~Acernese$^\text{5a,5c}$}\noaffiliation
\author{C.~Adams$^\text{6}$}\noaffiliation
\author{T.~Adams$^\text{7}$}\noaffiliation
% ...




\fake{}

\begin{abstract}
\noindent
We report a search for gravitational waves from the inspiral, merger 
and ringdown of binary black holes (BBH) with total mass between 25 and 
100 solar masses, in data taken at the LIGO and Virgo observatories 
between \ssixstart\ and \ssixend. 
The maximum sensitive distance of the detectors over this period for a 
(20,20)\Msun\ coalescence was \twentytwentyhorizon. 
No gravitational wave signals were found. We thus report upper limits on 
the astrophysical coalescence rates of BBH as a function of the component 
masses for non-spinning components, and also evaluate the dependence of 
the search sensitivity on component spins aligned with the orbital 
angular momentum. We find an upper limit at $90\%$ confidence on the 
coalescence rate of BBH with non-spinning components of mass between 19 
and 28$\,\Msun$ of \twentythreetwentythreelimit. 
\end{abstract}

\maketitle

\acrodef{GW}{gravitational-wave}
\acrodef{BBH}{Binary black hole}
\acrodef{PN}{post-Newtonian}
\acrodef{SNR}{signal-to-noise ratio}
\acrodef{LIGO}{Laser Interferometer Gravitational-wave Observatory}
\acrodef{LHO}{LIGO Hanford Observatory}
\acrodef{LLO}{LIGO Livingston Observatory}
\acrodef{LSC}{LIGO Scientific Collaboration}
\acrodef{CBC}{compact binary coalescence}
\acrodef{FAR}{false alarm rate}
\acrodef{VSR1}{the first Virgo science run}
\acrodef{VSR2}{Virgo's second science run}
\acrodef{VSR3}{third science run}
\acrodef{S5}{LIGO's fifth science run}
\acrodef{S6}{LIGO's sixth science run}

\section{Overview}
\noindent 
\ac{BBH} systems are a major class of possible \ac{GW} sources accessible 
to ground-based interferometric detectors such as 
LIGO~\cite{Abbott:2007kv} and Virgo~\cite{Accadia:2012zz}. 
As described in \cite{Collaboration:S5HighMass}, for higher-mass \ac{BBH}
systems, the merger and ringdown stages of the coalescence come into the 
detectors' sensitive frequency range and the search sensitivity is improved
by using inspiral-merger-ringdown (IMR) matched filter templates. Such 
templates were used in \cite{Collaboration:S5HighMass} to search for 
\acp{CBC} signals with total masses between $25$ and $100\,\Msun$ in LIGO
data. 


Our knowledge of possible high-mass \ac{BBH} source systems \cite{ratesdoc} is 
based on a combination of observations and astrophysical modelling: a summary 
of the recent evidence on both fronts is provided in 
\cite{Collaboration:S5HighMass}. A number of indicators point to the 
possibility of forming binary black holes with component masses $m_1$, $m_2$ of
 $\sim 20-30\,\Msun$ and beyond: in particular, predictions of the future fate 
of the high-mass Wolf-Rayet X-ray binaries IC10 X-1 and NGC 300 X-1 
\cite{Bulik:2008}; analyses of dynamical \ac{BBH} formation in dense stellar 
environments; the growing evidence for the existence of intermediate-mass black 
holes (e.g.~\cite{Davis:2011}); and population-synthesis modeling of 
low-metallicity environments \cite{Belczynski:2010}. 
A recent population-synthesis study \cite{Dominik:2012} that considered a wide 
range of astrophysical models found that in low-metallicity environments or 
under the assumption of weak wind-driven mass loss rates, the distribution of 
\ac{BBH} chirp masses $\mathcal{M} \equiv (m_1m_2)^{3/5}(m_1+m_2)^{-1/5}$ 
extended above $30\,\Msun$; for comparison, the chirp mass of a binary with 
component masses $(50,50)\,\Msun$ 
is $43.5\,\Msun$.  

In this paper we report a search for \ac{GW} signals from coalescence
of binary black holes with non-spinning components having masses $m_1,\,m_2$
between 1 and 99\,\Msun\ and total mass $M\equiv m_1+m_2$ between 25 and 
100\,\Msun, over the most recently taken coincident data from the LIGO and Virgo
observatories. A companion paper \cite{Collaboration:S6CBClowmass} describes
a search for low-mass binary inspiral signals with $2\leq M/\Msun \leq 25$ 
over these data, while a search of previous LIGO and Virgo data for \ac{BBH}
mergers signals with total mass $100$--$450\,\Msun$ is reported in \cite{S5IMBH}.

The joint science run used in this work---LIGO's sixth science run (S6) and 
Virgo's second (VSR2) and third (VSR3) science runs---was the most sensitive to 
date to signals from coalescing \ac{BBH}; this search was
also the first for high-mass BBH coalescences in Virgo data. We describe the
detectors and the joint S6-VSR2/VSR3 science run in Section~\ref{sec:data}. 
The search pipeline used here is similar to that of \cite{ihopePaper:2012,Collaboration:S5HighMass}, 
with changes to the ranking of events to account for variability of the noise 
background over the parameter space of the search and between detectors. We 
give a brief overview of the pipeline and describe changes relative to previous 
searches in Section~\ref{sec:pipeline}. 

The output of the analysis is a set of \emph{coincident events} where a 
potentially significant signal was seen in two or more detectors with 
consistent coalescence times and mass parameters. Events occurring at times 
when the detectors' environmental or instrumental monitor channels indicated a 
problem likely to corrupt the data are \emph{vetoed}: either removed from the 
search or placed in a separate category, depending on the severity of the 
problem. The significance of each remaining candidate event is measured by its 
\ac{FAR}, the expected rate of noise events with a detection statistic value 
(defined in Section~\ref{sec:pipeline}) at least as large as the candidate's.

As in previous LIGO-Virgo searches, the distribution of noise events in 
non-Gaussian data is estimated by applying unphysical time-shifts to data from 
different detectors. Events with low estimated \ac{FAR} are subject to a 
detailed followup procedure to check the consistency of the detector outputs 
around the event time and determine whether environmental disturbances or 
detector malfunction could have caused a spurious signal at that time. The 
search did not find any significant gravitational-wave candidate events; we 
describe the most significant events in Section~\ref{sec:results}.

We then evaluated the sensitivity of the search to coalescing \ac{BBH} at 
astrophysical distances by analyzing a large number of simulated signals 
(``injections'') added to the detector data. These are used to estimate the 
sensitivity of the search in terms of the sensitive distance in Mpc within 
which we would be able to detect a signal, averaged over the observation
time and over source sky location and orientation,
with significance above that of the loudest event observed in the search.
From this we set upper limits on the 
rate of such coalescences as a function of their component masses. For this 
purpose we used two recently-developed families of IMR waveforms. The improved 
EOBNRv2 family \cite{Pan:2011gk} was used to assess the sensitive range of the 
search for comparison with the previous high-mass \ac{BBH} search 
and to set upper limits on astrophysical 
coalescence rates; 
the {IMRPhenomB} waveform family \cite{Ajith:2009bn} was used to assess the 
sensitivity of the search to coalescences of \ac{BBH} with spinning components, 
where the component spins are aligned with the orbital angular momentum and 
thus the system does not precess.
We describe the injections performed and the resulting sensitivity distances 
and upper limits in Section~\ref{sec:UL}. 

To conclude, we briefly discuss outstanding issues for high-mass \ac{BBH} 
searches and prospects for the advanced detector era in 
Section~\ref{sec:discussion}.

\section{S6 and VSR2/VSR3 observations}
\label{sec:data}
\noindent
The US-based \ac{LIGO} comprises two sites: Hanford, WA and 
Livingston, LA. The data used in this search were taken during \ac{S6}, which 
took place between \ssixstart\ and \ssixend.  During \ac{S6} each of these 
sites operated a single 4-km laser interferometer, denoted as H1 and L1 
respectively. The 2-km H2 instrument at the Hanford site which operated in 
earlier science runs was not operational in \ac{S6}. Following \ac{S5} 
\cite{Abbott:2007kv}, several hardware changes were made to the \ac{LIGO} 
detectors in order to install and test prototypes of Advanced LIGO
\cite{0264-9381-27-8-084006} technology.
These changes included the installation of higher-powered
lasers, and the implementation of a DC readout system that included a new 
output mode cleaner on an Advanced LIGO seismic isolation 
table~\cite{Adhikari:2006}. In addition, the hydraulic seismic isolation 
systems 
were improved by fine-tuning their feed-forward paths.

The Virgo detector (denoted V1) is a single, 3-km laser interferometer located 
in Cascina, Italy.  The data used in this search were taken from both 
\ac{VSR2}~\cite{VirgoS2}, which ran from July 7, 2009 to January 11, 2010, and 
\ac{VSR3}, which ran from August 11, 2010 to October 20, 2010. 
In the period between \ac{VSR1} and 
\ac{VSR2}, several enhancements were made to the Virgo detector: a more 
powerful laser and a thermal compensation system were installed, and noise due 
to scattered light in the output beams was studied and mitigated.  During early 
2010, monolithic suspensions were installed, which involved replacing Virgo's 
test masses with new mirrors hung from fused-silica fibers.  VSR3 followed this 
upgrade.

A measure of the sensitivity of a detector to gravitational waves is its noise 
power spectral density (PSD) over frequency; typical PSDs for the S6 and VSR2/3 
science runs were given in \cite{Collaboration:S6CBClowmass}. Due to the improved 
low-frequency sensitivity of Virgo in VSR2/3~\cite{VirgoSuperattenuator}, 
the lower frequency cutoff of our analysis for V1 data was reduced to $30$\,Hz, 
compared to $40$\,Hz for LIGO data. 

\section{Data Analysis Pipeline}
\label{sec:pipeline}
\noindent
Our search algorithm, which was described in detail in \cite{ihopePaper:2012,
Collaboration:S5HighMass}, is based on matched filtering the data in each 
detector against a template bank of IMR waveforms, recording local maxima of 
\ac{SNR} as \emph{triggers}, then testing these triggers for consistency of 
coalescence time and mass parameters between two or more detectors via a 
coincidence test~\cite{Robinson:2008}, and for their consistency with the 
template waveform via the $\chi^2$ test~\cite{Allen:2004}. The $\chi^2$ 
test is necessary to suppress noise transients (see Section~\ref{sec:DQ}), which 
cause a much larger rate of triggers with high SNR than expected in Gaussian 
noise. 

\begin{figure}[tp]
\vspace*{-0.3cm}
\hspace*{0.0cm}
\includegraphics[width=0.85\columnwidth]{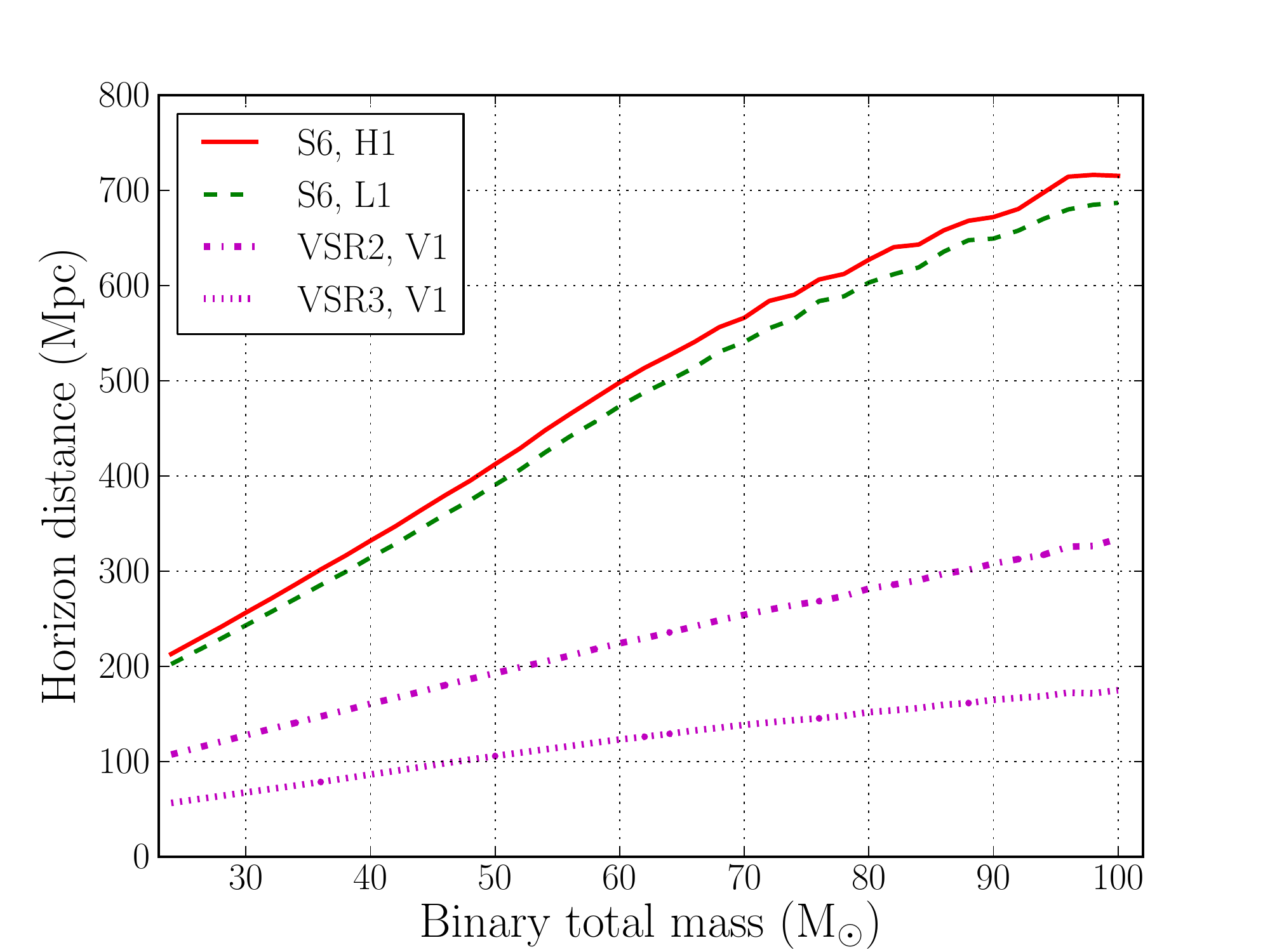}
\caption{Horizon distances for non-spinning equal-mass IMR signals in the LIGO and
Virgo detectors, using EOBNRv2 as signal model, averaged over periods of data
when the detector sensitivities were near optimal for S6 and VSR2 and -3,
respectively.}
\label{fig:horizons}
\end{figure}

\subsection{Filter templates and optimal search sensitivity}

As filter templates we used the same family of waveforms as described in 
\cite{Collaboration:S5HighMass} constructed using the results of 
\cite{Buonanno:2007pf}, which we will refer to as EOBNRv1. The parameter space  
covered by our templates was also unchanged, ranging from 1 to $99\,\Msun$ for
the binary component masses $m_1$, $m_2$, and from 25 to $100\,\Msun$ for the 
total binary mass $M = m_1+m_2$; a search covering total mass values between 
2 and $25\,\Msun$ was reported in \cite{Collaboration:S6CBClowmass}.

The recently-implemented EOBNRv2 \cite{Pan:2011gk} and IMRPhenomB 
\cite{Ajith:2009bn} waveforms are more accurate than their predecessors in that 
they are better fits to waveforms produced by numerical relativity (NR) 
simulations, and because the NR waveforms themselves have improved in accuracy 
and cover a wider range of parameter space \cite{Hannam:2007ik,Hannam:2010ec}, 
\cite{Buchman:2012dw,SpECwebsite}. We discuss the relevant properties of the 
EOBNRv2 and IMRPhenomB waveform models in Sections \ref{sec:EOBNRrates} and 
\ref{sec:Phenomrates}. 

We investigated whether these 
improved waveform families could be efficiently detected by a bank of 
EOBNRv1 filter templates. The method used was to calculate the overlap between 
an EOBNRv2 or IMRPhenomB signal waveform and an EOBNRv1 template, maximizing 
over the parameters of the EOBNRv1 template (see 
\cite{Damour:1998zb} for the general method).
For EOBNRv2 signals, over the range of mass ratio $1\leq q \equiv m1/m2 \leq 6$ 
and total mass $25\leq M/\Msun \leq 100$, in the worst case the maximized overlap 
(effectualness) of EOBNRv1 templates was greater than 0.97. For non-spinning 
IMRPhenomB signals, the smallest effectualness was greater than 0.98. Thus, the 
use of EOBNRv1 templates did not significantly degrade the efficiency of our 
search, for non-spinning signals. The more recent waveform families are, however, 
useful in more accurately determining the sensitivity of the search to 
astrophysical BBH mergers.

As a simple measure of the maximum possible search sensitivity, we show in 
Figure~\ref{fig:horizons} the horizon distances for equal-mass EOBNRv2 signals, 
defined as the distances at which an optimally-oriented coalescence 
directly overhead from a given detector would have an expected SNR of 8 for an 
optimal matched filter; as in \cite{Collaboration:S6CBClowmass} we average 
these distances over periods of data for which 
the detector sensitivities were optimal, or close to optimal, over each
science run. We see that the maximum sensitive distance for a (20,20)\Msun\ 
coalescence in each of the LIGO detectors was approximately \twentytwentyhorizon.
The detector sensitivities varied significantly over the observation time of 
this search, as detailed in \cite{Collaboration:2012wu}. Sensitive distances 
averaged over observation time and over source sky location and orientation 
are reported in Section~\ref{sec:EOBNRrates}.

\begin{figure*}[htp]
\hspace*{-5mm}
\includegraphics[width=0.85\columnwidth]{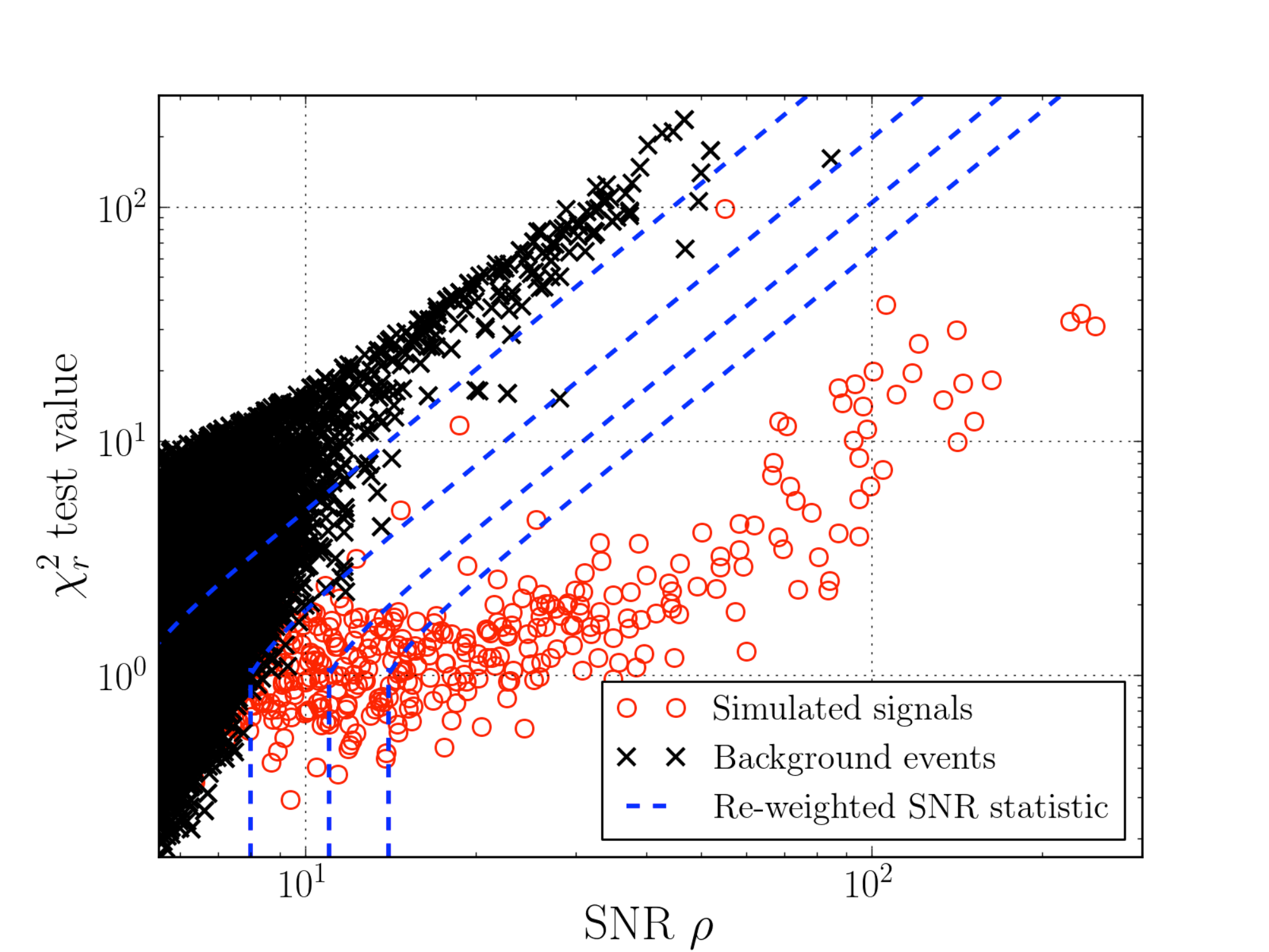}
\hspace*{8mm}
\includegraphics[width=0.85\columnwidth]{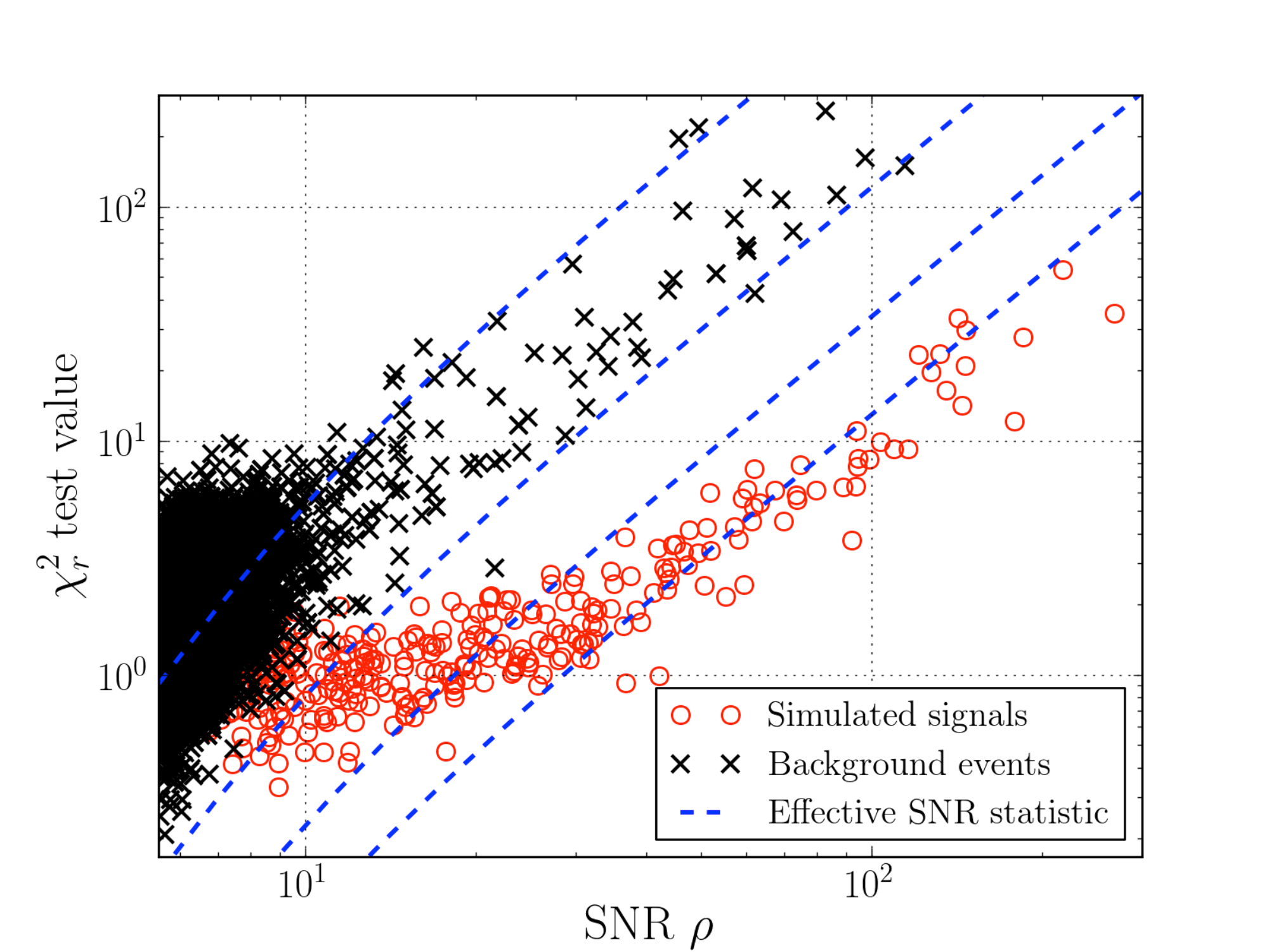}
\caption{
Representative distributions of SNR and $\chi^2_r$ values for simulated
signal (red circle) and background (black `$\times$') triggers in the LIGO 
detectors, with contours of the detection statistics used in the search. Note the 
systematically lower values of $\chi^2$ for background events with SNR $\rho>10$
in shorter-duration templates (right plot) compared to longer-duration (left plot). 
\emph{Left}---triggers with template duration greater than 0.2\,s; dashed lines
indicate contours of constant re-weighted SNR statistic, Eq.~\eqref{eq:new_snr}.
\emph{Right}---triggers with template
duration below 0.2\,s; dashed lines indicate contours of constant effective SNR,
Eq.~\eqref{eq:effsnr}.}
\label{fig:snr_chisq_duration}
\end{figure*}
\subsection{Background estimate and event ranking statistic}
\label{sec:highstat}
After obtaining a list of candidate coincident events, each consisting of two or 
more triggers with consistent template masses and coalescence times, we estimate 
the significance of each event relative to background.  Our background distribution 
is obtained by finding coincident events after applying unphysical time-shifts 
(greater than light travel time) to data from different detectors; we performed 100 
time-shifted analyses using multiples of a $(0,5,10)$\,s time offset for (H1,L1,V1) 
data, respectively. In order to compare the coincident events to background we 
require a ranking statistic.  

The aim of our ranking statistic is to optimize the separation of signal from 
background in the search. We tuned it by studying the distributions of triggers 
over SNR ($\rho$) and $\chi^2$ for time-shifted background events, and for 
events resulting from simulated IMR signals added to the data (``software 
injections''). We found that the distribution of background triggers depended 
strongly on \emph{template duration}, a trigger parameter determined by the 
binary component masses and by the lower frequency cutoff used in the analysis,
which was taken to be $40$\,Hz for LIGO data and $30$\,Hz for Virgo. 
Template durations varied between approximately $0.05$\,s for templates with the 
highest total mass $100\,\Msun$, in the LIGO detectors, to several seconds for 
lower-mass signals. 
We compare signal-background separation for triggers from longer \emph{vs.}\
shorter-duration templates, in a representative period of LIGO data, in
Figure~\ref{fig:snr_chisq_duration}. 
The performance of the 
$\chi^2$ test was markedly worse for templates shorter than $0.2$\,s in LIGO and 
VSR3 data, with some noise triggers in these templates having large SNR but a 
relatively small $\chi^2$ value, comparable to simulated IMR signals. For 
signals seen in LIGO, this threshold value of $0.2\,$s corresponds to a total 
mass of approximately $45\,\Msun$ for equal-mass systems, or a total mass of 
approximately $90\,\Msun$ for the most asymmetric templates used in the search. 
%

The poor performance of the $\chi^2$ test for short templates in LIGO and VSR3 
data can be attributed to the small number of template cycles 
over which SNR is accumulated in the templates. By contrast, the $\chi^2$ test was 
found to be effective in penalizing noise artefacts in VSR2 data over the entire 
parameter space of the search. 

We divided coincident events into two bins: one for which all participating triggers 
from H1 or L1 (or V1, in VSR3 data) had template durations above $0.2$\,s (``long 
duration events'') and one where at least one trigger from H1 or L1 (or V1, in VSR3) 
had a template duration below $0.2$\,s (``short duration events''). 

Due to the different distributions of background triggers over SNR and $\chi^2$
for longer-duration \emph{vs.}\ shorter-duration templates, as illustrated in 
Figure~\ref{fig:snr_chisq_duration}, we found that a 
different choice of ranking statistic was appropriate for the two bins. 
For all triggers participating in long duration events, and all V1 triggers in VSR2 
data, we used the re-weighted SNR $\hat{\rho}$ statistic 
\cite{Collaboration:S6CBClowmass} defined as
\begin{equation}\label{eq:new_snr}
 \hat{\rho} = \begin{cases} 
 {\displaystyle \frac{\rho}{[(1+(\chi^2_r)^3)/2]^{1/6}}} & \mbox{for } \chi^2_r > 1, \\
 \rho & \mbox{for } \chi^2_r \leq 1,
 \end{cases}  
\end{equation}
where $\chi^2_r \equiv \chi^2/(2p-2)$, and where we chose the number of frequency 
intervals $p$ used in the evaluation of $\chi^2$~\cite{Allen:2004} to be 
$10$~\cite{Collaboration:S5HighMass}. 
For H1 or L1 triggers, or V1 triggers in VSR3 data, participating in short duration
events we used the effective SNR statistic of \cite{Collaboration:S5HighMass}:
\begin{equation} \label{eq:effsnr}
 \rho_{\rm eff} = \frac{\rho}{[\chi^2_r(1+\rho^2/50)]^{1/4}} .
\end{equation}
The detection statistic, ``combined SNR'' $\rho_c$, is then given by the quadrature 
sum of single-detector statistics, over the coincident triggers participating in an 
event.\footnote{Note that, as in previous searches, we applied a weak 
$\rho$-de\-pendent cut on the $\chi^2_r$ values, to remove triggers with very low 
statistic values: the effect of this cut may be seen in 
Figure~\ref{fig:snr_chisq_duration} where the top left of each plot is empty.}

We calculate the detection statistic values separately in different coincident 
times (times when two or more detectors are recording data, labelled by the active 
detectors), due to their different background event distributions and different 
sensitivity to astrophysical signals. The \acp{FAR} of coincident events are 
estimated by first comparing their $\rho_c$ values to those of time-shifted 
background events in the same bin by duration, and with the same event type 
(\textit{i.e.}\ the same detectors participating in coincidence). 
The final detection statistic of the search, combined FAR, is determined by 
ranking the event's FAR against the total distribution of background FAR values, 
summed over both bins in template duration and over all event types within each 
coincident time: see Eq.~(III.7-8) of \cite{Collaboration:S5HighMass}. 

\subsection{Data quality vetoes} 
\label{sec:DQ} 
The gravitational-wave strain data from the detectors contains a larger number of 
transient noise events (glitches) with high amplitude than would occur in colored 
Gaussian noise. In order to diagnose and remove these transients, each of the 
\ac{LIGO} and Virgo observatories is equipped with a system of environmental and 
instrumental monitors that have a negligible sensitivity to gravitational waves but 
may be sensitive to glitch sources. These sensors were used to identify 
times when the detector output was potentially corrupted~\cite{Slutsky:2010ff,Christensen:2010,
Robinet:2010zz,SeisVeto}. We grouped these times into two categories: periods 
with strong and well-understood couplings between non-GW transient noise 
sources and detector output, and periods when a statistical correlation was found 
although a coupling mechanism was not identified. In our primary search, both for 
the identification of GW candidates and the calculation of upper limits, times in 
both these categories, and any coincident events falling in these times, were 
removed (``vetoed'') from the analysis. We also performed a secondary search for 
possible loud candidate events, in which only times with clear coupling of non-GW 
transients to detector output were vetoed. The total time searched for GW 
candidate events, in which only the first category of vetoes were applied, 
was \totalLoudTime. 

Even after applying vetoes based on auxiliary (environmental and instrumental) 
sensors, significant numbers of delta-function-like glitches with large amplitude 
remained unvetoed in the LIGO detectors. It was found that these caused artifacts 
in the matched filter output over a short time surrounding the glitch: thus, 8\,s 
of time on either side of any matched filter SNR exceeding 250 was additionally 
vetoed. Times removed from the primary search by this veto were still examined for 
loud candidate events. 

Approximately \totalpostVeto\ of coincident search time remained after applying all 
vetoes. 
Additionally, approximately $10\%$ of the data, designated {\it playground}, was 
used for tuning and data quality investigations.  These data were searched for
gravitational waves, but not used in calculating upper limits.  After all vetoes 
were applied and playground time excluded, there was \HLVlt~of H1L1V1 coincident 
time, \HLlt~of H1L1 time, \HVlt~of H1V1 time, and \LVlt~of L1V1 time, giving a 
total analysis time of \totalTime.

\section{Search results}
\label{sec:results}
\begin{figure}[tp]
\includegraphics[width=0.9\columnwidth]{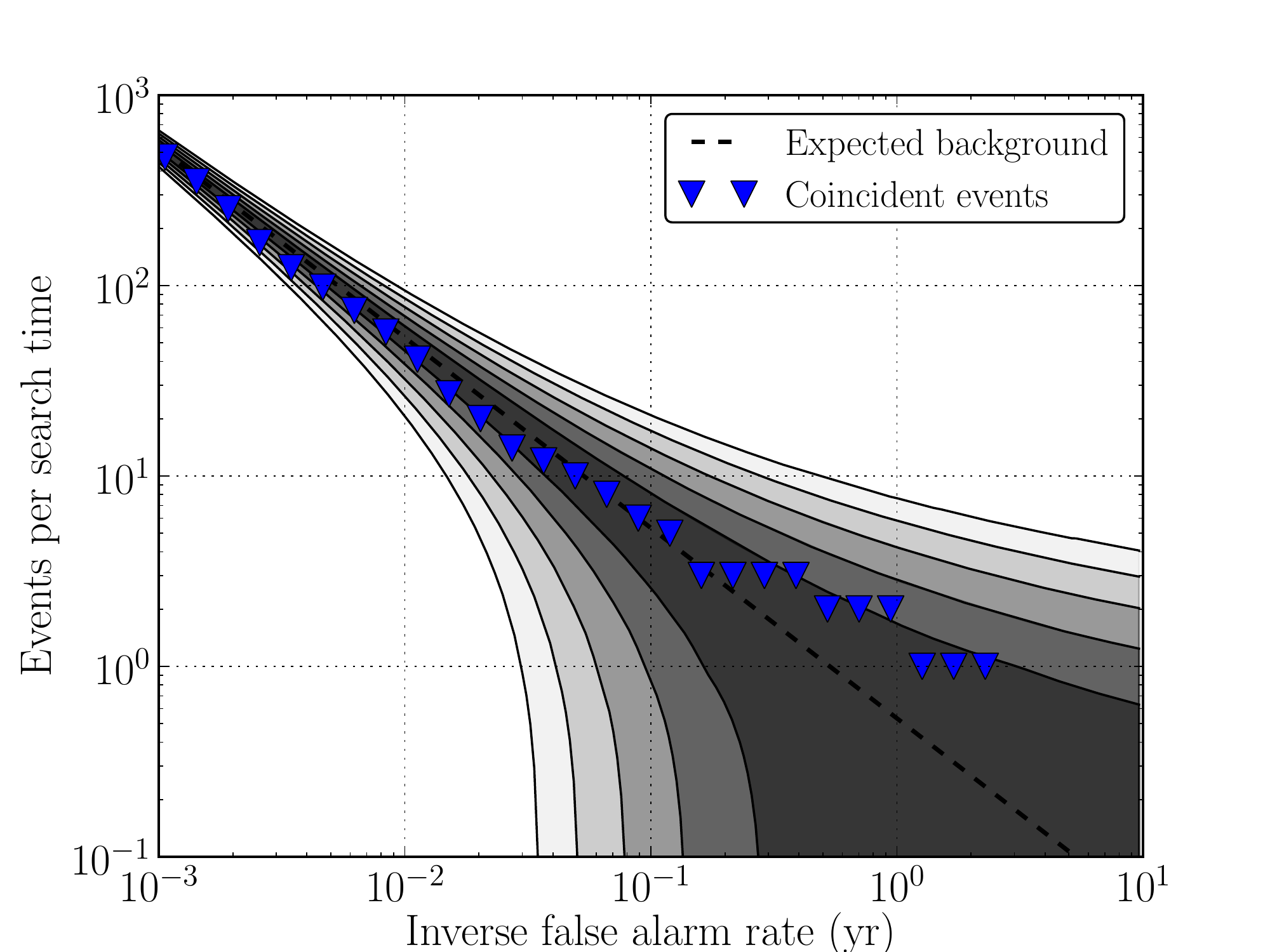}
\caption{Cumulative distribution of coincident events found in the search 
\emph{vs.}\ estimated inverse false alarm rate (FAR), over the total time 
searched for possible GW candidates, \totalLoudTime. 
Grey contour shading indicates the consistency at $1\sigma$ (dark) through 
$5\sigma$ (light) level of search coincident events with the expected 
background. 
}
\label{fig:IFAR_summary}
\end{figure}
\begin{figure*}[tp]
\includegraphics[width=0.9\columnwidth]{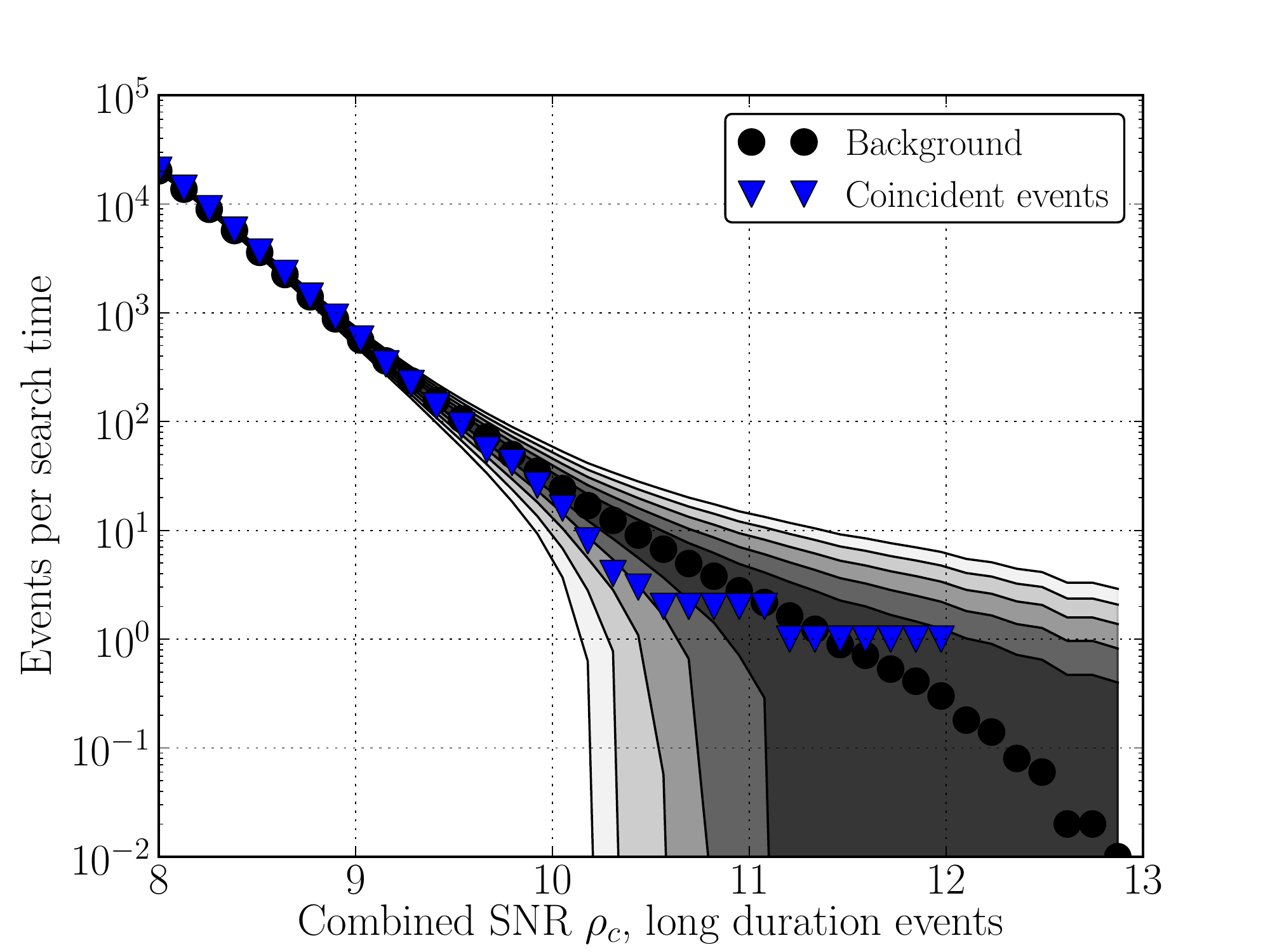}
\includegraphics[width=0.9\columnwidth]{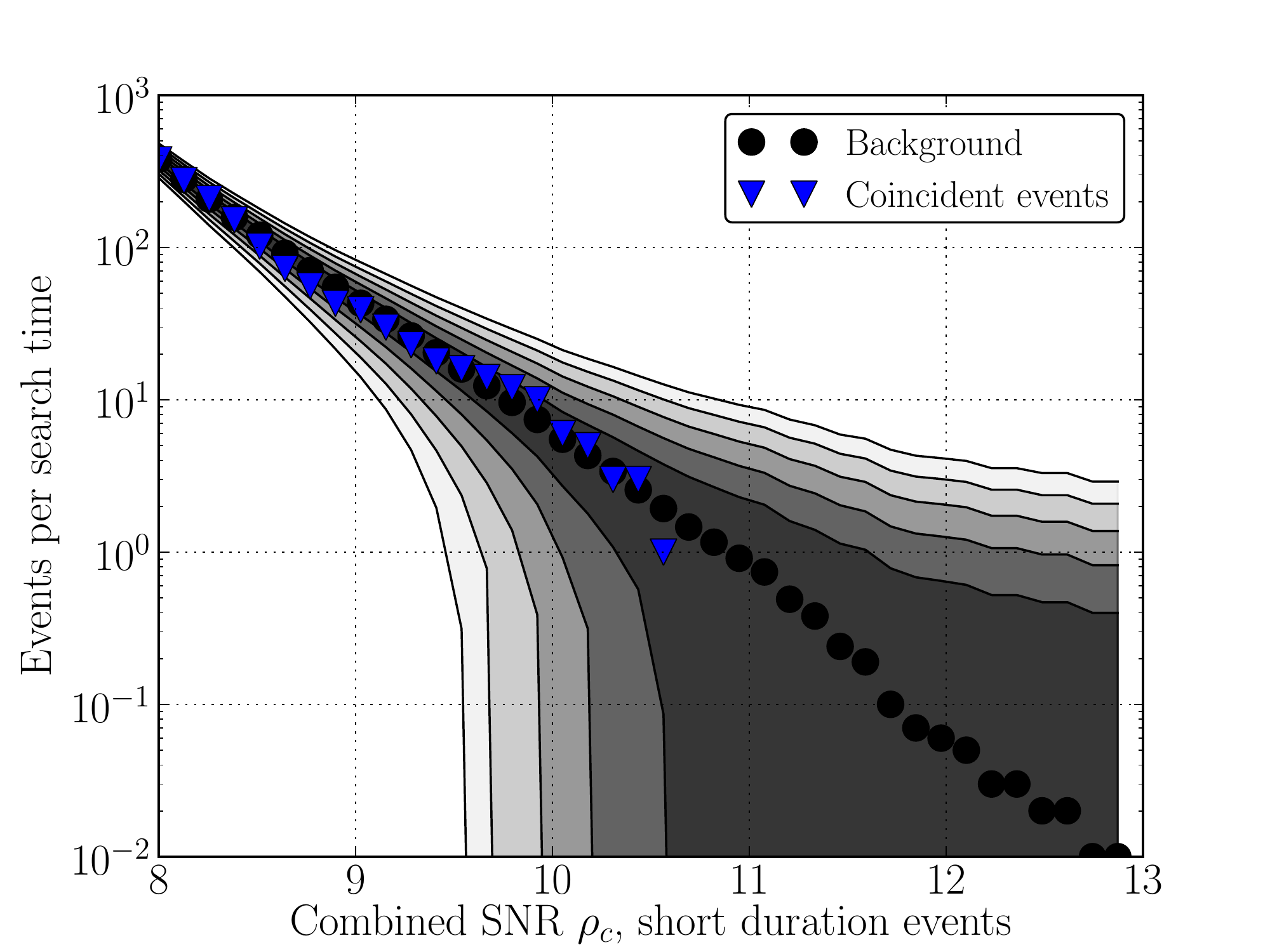}
\caption{Cumulative distributions of coincident events and estimated background
over combined SNR statistic $\rho_c$, over the total time searched for possible GW 
candidates. Grey shaded bands indicate $1\sigma$--$5\sigma$ consistency with the 
estimated background distribution. 
\emph{Left}---Distribution for ``long duration'' events. 
\emph{Right}---Distribution for ``short duration'' events. The two event bins and the
combined SNR statistic are described in Section~\ref{sec:highstat}.
\label{fig:SNR_summary}
}
\end{figure*}
\noindent
We found no significant or plausible gravitational-wave detection candidates 
above the noise background of the search. The cumulative distribution of 
coincident events found in the search \emph{vs}.\ estimated inverse false 
alarm rate (FAR) is shown in Figure~\ref{fig:IFAR_summary}. The distribution 
is consistent with the expected background over the total time searched for 
GW candidates, \totalLoudTime.

The most significant coincident event found in the search, with lowest 
estimated \ac{FAR} (highest inverse \ac{FAR}), 
was at GPS time \loudestGPS\ and had an estimated \ac{FAR} of \loudestFAR. 
This event, an H1V1 coincidence in H1V1 coincident time with SNR values of 31 in 
H1 and 5.50 in V1 and a combined SNR statistic value of $\rho_c = \loudestRhoc$, 
occurred at a time when several short-duration, non-Gaussian transient excess 
power events were visible, over a period of several seconds, in time-frequency 
maps of the GW strain channel in H1. The high trigger SNR in H1 was caused by 
the first of these transients; however the behavior of the strain and SNR time 
series over the following few seconds is strongly inconsistent with a high-mass 
binary coalescence signal. The $\chi^2$ test value in H1 was high (approximately 
190), but not sufficiently high to rule out the trigger as a candidate. 

The frequency spectrum of the noise transients in H1 indicates their probable 
origin as stray light scattered into the interferometer beam. 
The appearance of the event in V1 was consistent with both a quiet GW signal and 
with random noise. 
The event time was not vetoed in the search, as none of the instrumental or 
environmental channels at H1 which we found to be significantly correlated with 
noise transients in the GW channel showed a malfunction at the event time. 
 
The second loudest event was at GPS \secondloudestGPS\ with an estimated \ac{FAR} 
of \secondFAR. It was an H1L1 coincidence in H1L1 time with SNRs of 13 in H1 and 70 
in L1 and a combined SNR statistic value of $\rho_c = \secondRhoc$. 
This time was subject to a veto in H1, due to a problem in a high-voltage 
power supply near the output photodiodes (affecting some weeks of data) which 
caused bursts of broad-band non-Gaussian noise. However, since this excess noise 
was not sufficiently severe to preclude detection, the time was still searched for 
possible high-SNR candidate events, as described in Section~\ref{sec:DQ}. 
The trigger in L1 was caused by a high amplitude non-Gaussian transient of very 
short duration, part of a population of sporadic glitches for which no effective 
veto could be found. This event also failed a detailed followup, as time-frequency 
maps of excess power, and the time series of SNR and $\chi^2$ in H1, were 
inconsistent with an IMR signal. 

The next few most significant events had estimated \acp{FAR} of a few per year and
were thus entirely consistent with background. 

We show the $\rho_c$ 
distributions of coincident events and time-shifted background events, for the
total time searched for possible GW candidates, in Figure~\ref{fig:SNR_summary}, 
separating the long- and short-duration events since $\rho_c$ is a different 
function of $\rho$ and $\chi^2$ in each. 
The $\rho_c$ values of all the loudest search events were less than $12$ (for 
comparison, a BBH coalescence signal with an \ac{SNR} of 8 in each of two 
detectors would give approximately $\rho_c\simeq 11.3$). 
Thus, the data quality veto procedure, in combination with the signal-based 
$\chi^2$ test, 
were sufficient to remove or suppress loud detector artefacts to a level where 
the sensitivity of our search is not greatly impaired.

\section{Upper Limits on BBH Coalescence Rates}
\label{sec:UL}
\noindent
Given the null result of our search for \ac{BBH} coalescence signals, we wish to 
set observational limits on the astrophysical rates of such signals, over 
regions of parameter space where our search has non-negligible sensitivity. As 
discussed in \cite{Collaboration:S5HighMass}, the distance reach of this 
high-mass search is such that the source population can be treated as 
approximately homogeneous over spatial volume; thus we aim to set limits on the 
rate density of coalescences, in units of \perMpcyr.

\subsection{Upper limit calculation procedure}

Our upper limit calculation is similar to that performed in 
\cite{Collaboration:S6CBClowmass}: it is based on the loudest event 
statistic \cite{Biswas:2007ni} applied as described in Section V of 
\cite{Collaboration:S5HighMass} with minor improvements in implementation. 
We divide the data into 9 periods of approximately 6 weeks each; in 4 of these
only H1L1 coincident time was recorded, whereas in the remaining 5 we had
four types of coincident time (H1L1, H1V1, L1V1 and H1L1V1). 

In each of the resulting 24 analysis times, we estimate the volume to which 
the search is sensitive by reanalyzing the data with the addition of a large 
number of simulated signals (``software injections'') in order to model the 
source population. Our ability to detect a signal depends upon the parameters 
of the source, including the component masses and spins (magnitudes and 
directions), the distance to the binary, its sky location, and its orientation 
with respect to the detectors. Numerous signals with randomly chosen parameter 
values were therefore injected into the data. 

To compute the sensitive volume over a given range of binary masses (``mass 
bin''), we perform a Monte Carlo integration over the other parameters to 
obtain the efficiency of the search---determined by the fraction of simulated 
signals found louder than the loudest observed coincident event in each analysis 
time---as a function of distance. Integrating the efficiency over distance then 
gives the sensitive volume for that analysis time, and an associated sensitive 
distance.

We then estimate the likelihood parameter $\Lambda$ of signal \textit{vs.}\ 
background at the combined FAR value of the loudest observed event in each 
analysis time, for each mass bin, as described in \cite{Biswas:2007ni,
Collaboration:S5HighMass}. 
Using these $\Lambda$ values and the estimated sensitive volumes we find the 
probability of the measured loudest FAR value as a function of the astrophysical 
merger rate, \textit{i.e.}\ the likelihood of the data given the model, in each 
analysis time and for each mass bin. 
Given a prior probability distribution over the rate, in 
each mass bin, we then multiply by the likelihoods of the loudest events from 
all the analysis times to form a posterior over rate: see 
\cite{Collaboration:S5HighMass} (Section V and Erratum) for relevant formulae. 

The likelihood function 
for each analysis time depends 
on the sensitive volume$\times$time searched; 
however, the sensitive volumes have statistical uncertainties due to the 
finite number of injections performed in each mass bin, 
and systematic uncertainties 
in the amplitude calibration for each detector. As detailed in 
\cite{Collaboration:S6CBClowmass}, we take an overall $42\%$ uncertainty in 
volume due to calibration errors. We marginalize over statistical 
uncertainties for each analysis time separately, but since systematic 
calibration errors may be significantly correlated between analysis times we 
perform this marginalization \cite{Fairhurst:2007qj} once after combining the 
likelihoods from all analysis times. We then find the 90\% confidence upper 
limit based on the marginalized posterior distributions over rate. 

We also calculated an average sensitive distance in each mass bin, defined as
the radius of a sphere such that the sphere's volume, multiplied by the total
search time, equals the total sensitive volume$\times$time over all analysis
times.

Since the injected waveforms are phenomenological models, our upper limits
will also be systematically affected to the extent that the true IMR waveforms 
differ from these models. These uncertainties are difficult to quantify over 
the search as a whole and we will not attempt to incorporate them into our 
quoted limits. A comparison of EOBNRv2 waveforms against numerical relativity 
simulations for mass ratios $q=1,4,6$ shows possible SNR biases of at most a 
few percent within the total mass range $25\leq M/\Msun\leq 100$.

The rate priors that we use for non-spinning EOBNRv2 signals in the S6-VSR2/3 
search are derived from the results of the S5 high-mass \ac{BBH} search 
\cite{Collaboration:S5HighMass}. The original results from this search were 
affected by an incorrect treatment of marginalization over errors in the 
sensitive volume and flaws in the numerical procedure used to estimate the 
$\Lambda$ values, resulting overall in an over-conservative set of 90\% rate 
upper limits. These problems were recently addressed, leading to revised upper 
limits from S5 data (\cite{Collaboration:S5HighMass}, Erratum); the resulting 
revised posteriors over coalescence rate were used as priors for our main upper 
limit calculation. 
Revised rate upper limits from S5 data alone are also included in
Table~\ref{t:UpperLimits}. 
Note that, as priors \emph{for the S5 calculation,} uniform probability 
distributions over rate were taken; this uniform prior 
is a conservative choice for setting upper limits. 

\begin{figure*}[tp]
\hspace*{-2mm}
\includegraphics[width=0.9\columnwidth]{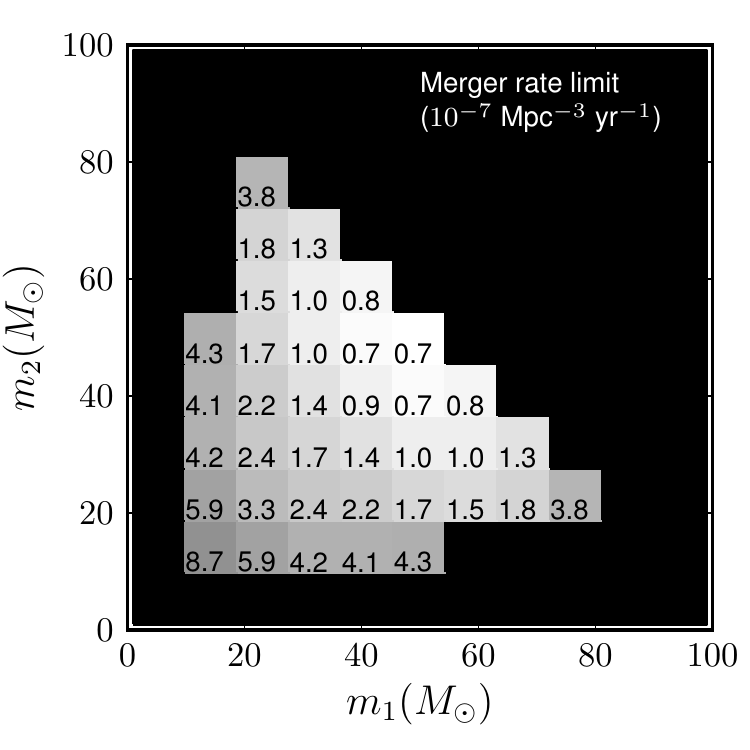}
\hspace*{4mm}
\includegraphics[width=0.9\columnwidth]{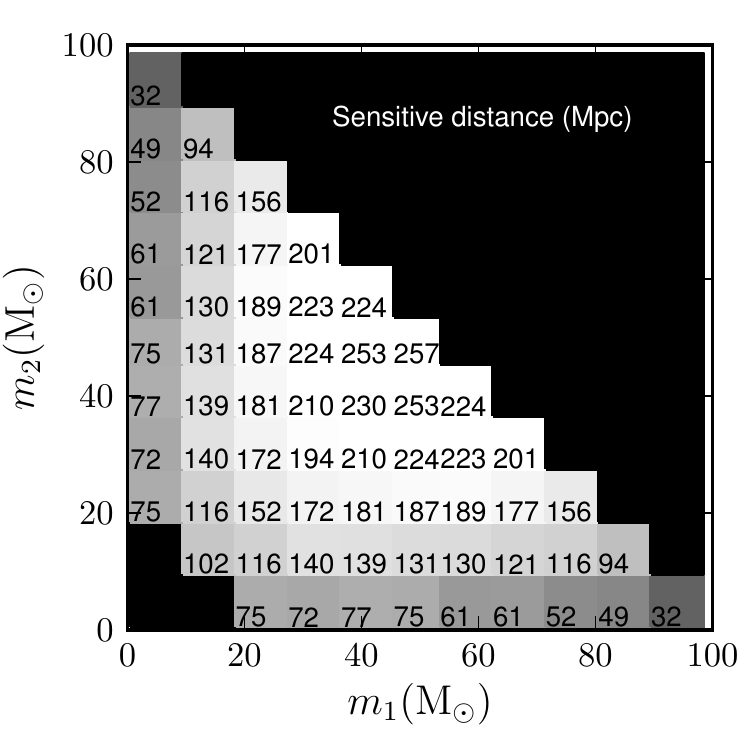}\\
\caption{\emph{Left}---Upper limits (90\% confidence) on BBH coalescence 
rates in units of $10^{-7}\perMpcyr$ as a function of binary component 
masses, evaluated using EOBNRv2 waveforms. 
\emph{Right}---Average sensitive distance for this search to binary 
systems described by EOBNRv2 signal waveforms, in Mpc.
}
\label{fig:EOBNRUL}
\end{figure*}

\begin{table*}[tp]
\caption{\label{t:UpperLimits}
Search sensitive distances and coalescence rate upper limits, quoted over
9\,\Msun-wide component mass bins labelled by their central values. We also
quote the chirp mass $\mathcal{M}$ at the centre of each bin.
 The sensitive distance in Mpc (averaged over the observation time and over
source sky location and orientation) is given for EOBNR waveforms in S5 data 
rescaled for consistency with NR results 
\cite{Collaboration:S5HighMass},
 and for EOBNRv2, IMRPhenomB non-spinning (``PhenomB nonspin'') and IMRPhenomB 
spinning (``PhenomB spin'') waveforms in the S6-VSR2/3 data. 
 The last two columns report 90\%-confidence rate upper limits in units of
$10^{-7}\,\perMpcyr$, for bins with component mass ratios $1\leq m_1/m_2\leq 
4$, for S5 data (revised relative to \cite{Collaboration:S5HighMass})
and the cumulative upper limits over S5 and S6-VSR2/3 data, as presented in
this work.
}
\begin{center}
\begin{tabular}{ccc|cccc|cc}
\hline
 & Waveforms  &  & EOBNR    & EOBNR & PhenomB nonspin & PhenomB spin & EOBNR & EOBNR \\
 & Search data & & { S5}        & {S6-VSR2/3} & {S6-VSR2/3}  & {S6-VSR2/3} & {S5} & {S5$\,+\,$S6-VSR2/3} \\
\hline
 $m_1$ & $m_2$ & $\mathcal{M}$ & Distance & Distance & Distance & Distance & Upper Limit & Upper Limit  \\
 { (\Msun)}  & { (\Msun)} & { (\Msun)} & { (Mpc)} & { (Mpc)} & { (Mpc)} & { (Mpc)} &
  ($10^{-7}\,\perMpcyr$)  &
  ($10^{-7}\,\perMpcyr$) \\
\hline \hline
14  &  14  &  13  &  81  &  102  &  105  &  106  &  18  &  8.7  \\
23  &  14  &  16  &  95  &  116  &  126  &  126  &  12  &  5.9  \\
32  &  14  &  18  &  102  &  140  &  132  &  135  &  8.8  &  4.2  \\
41  &  14  &  21  &  107  &  139  &  141  &  145  &  7.8  &  4.1  \\
50  &  14  &  22  &  107  &  131  &  137  &  149  &  8.2  &  4.3  \\
23  &  23  &  20  &  116  &  152  &  148  &  149  &  7.4  &  3.3  \\
32  &  23  &  24  &  133  &  172  &  172  &  179  &  4.9  &  2.4  \\
41  &  23  &  27  &  143  &  181  &  178  &  183  &  4.3  &  2.2  \\
50  &  23  &  29  &  145  &  187  &  188  &  198  &  3.4  &  1.7  \\
59  &  23  &  32  &  143  &  189  &  188  &  192  &  3.2  &  1.5  \\
68  &  23  &  34  &  140  &  177  &  180  &  191  &  3.7  &  1.8  \\
77  &  23  &  36  &  119  &  156  &  176  &  170  &  5.6  &  3.8  \\
32  &  32  &  28  &  148  &  194  &  190  &  197  &  3.4  &  1.7  \\
41  &  32  &  32  &  164  &  210  &  219  &  220  &  2.5  &  1.4  \\
50  &  32  &  35  &  177  &  224  &  221  &  214  &  1.9  &  1.0  \\
59  &  32  &  38  &  174  &  223  &  221  &  214  &  2.0  &  1.0  \\
68  &  32  &  40  &  162  &  201  &  199  &  210  &  2.4  &  1.3  \\
41  &  41  &  36  &  183  &  230  &  222  &  224  &  1.6  &  0.9  \\
50  &  41  &  39  &  191  &  253  &  253  &  258  &  1.4  &  0.7  \\
59  &  41  &  43  &  194  &  224  &  239  &  236  &  1.4  &  0.8  \\
50  &  50  &  44  &  192  &  257  &  218  &  217  &  1.4  &  0.7  \\
\hline
\end{tabular}
\end{center}
\end{table*}

\begin{figure}[tp]
\includegraphics[width=0.95\columnwidth]{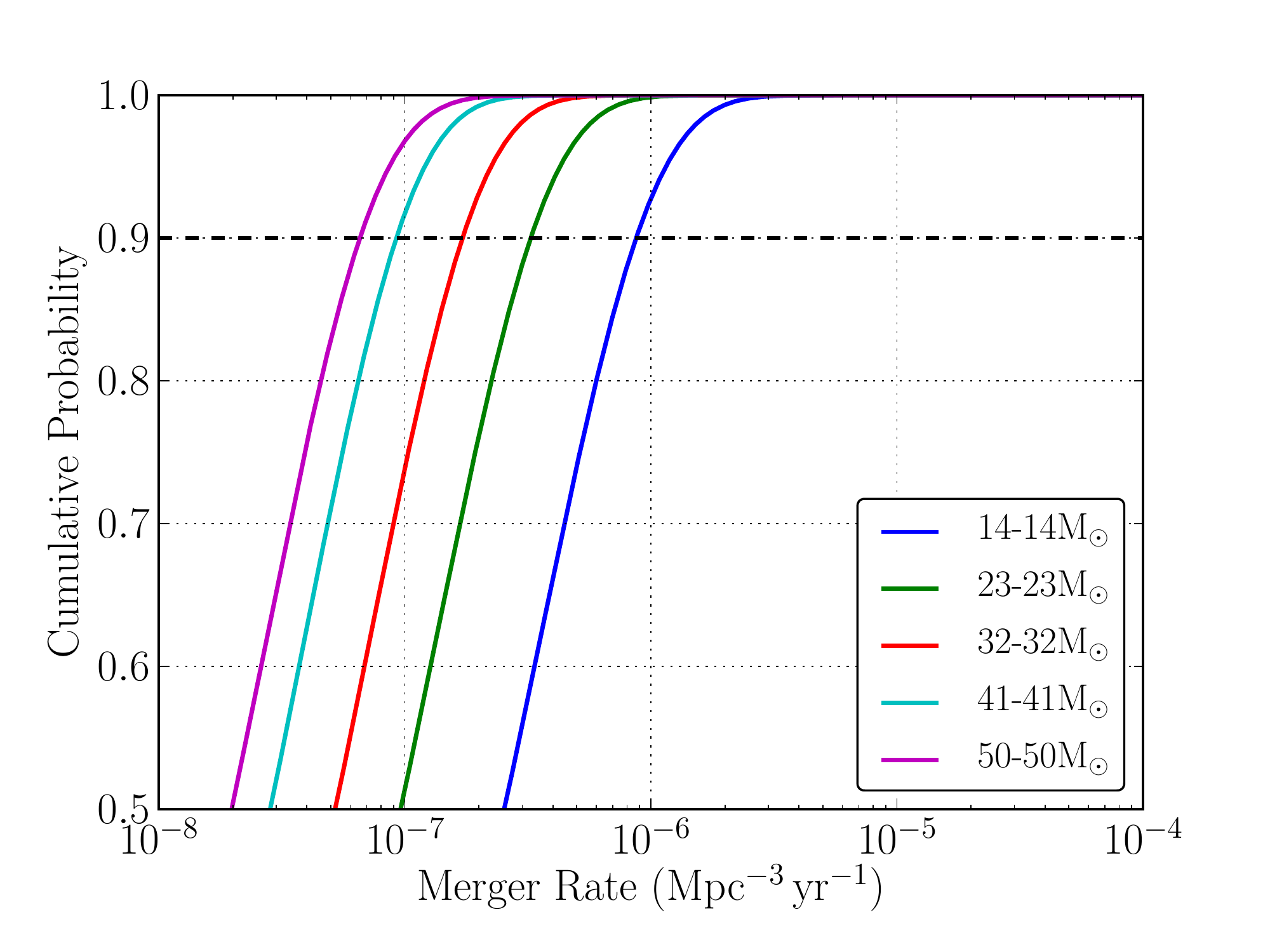}
\hspace*{3mm}
\caption{Cumulative posterior probabilities over astrophysical merger
rate, 
for the bins shown
in Figure~\ref{fig:EOBNRUL} with central values $m_1=m_2= 50$, 41, 32, 23, 
14$\,\Msun$ (left to right).
We show the probability level corresponding to the 90\% confidence rate 
limit (dashed horizontal line).
These posteriors were evaluated for signals described by the EOBNRv2
waveform family in S6 data using S5 search results as prior information.
\label{fig:ratepdfs}
}
\end{figure}

\subsection{Rate limits from EOBNRv2 injections}
\label{sec:EOBNRrates}

In order to evaluate the search sensitivity to non-spinning IMR signals over
a wide parameter space, and to allow a comparison to previous search results, 
we used an implementation of the recently developed EOBNRv2 waveform family 
\cite{Pan:2011gk} as simulated signals. 

The EOBNRv2 waveform family was designed using results from \cite{DIN, 
Damour2009a, Pan2010hz, Barausse:2011kb}. 
The inspiral waveforms in the EOBNRv2 model are improved over EOBNRv1 by 
calibrating two adjustable parameters against five highly accurate NR 
simulations of mass ratios $q=1,2,3,4$ and $6$, generated by the pseudospectral 
code SpEC~\cite{Buchman:2012dw,SpECwebsite}. These two parameters are the 
pseudo-4PN and 5PN coefficients $a_5$ and $a_6$ entering the EOB radial 
potential (see, \textit{e.g.}, Eq.~(II7a) in 
Ref.~\cite{Collaboration:S5HighMass} and related discussion.) 
EOBNRv2 also improves over EOBNRv1 by including all known \ac{PN} corrections in 
the amplitude, by using a more accurate estimate of the radiated energy flux, 
by dropping the assumption of quasi-circular orbits, by improving the matching 
of the inspiral-plunge waveform to the ringdown modes, and by improving the 
extrapolation to large mass ratios. 
The differences between these EOBNRv2 and NR waveforms are 
comparable with numerical errors in the NR simulations. 

EOBNRv2 injections were distributed to ``over-cover'' the parameter range of 
the search, in order to ensure complete coverage of the mass bins used in 
\cite{Collaboration:S5HighMass}. The injections were distributed approximately 
uniformly over the component masses $m_1$ and $m_2$, within the ranges 
$1\,\Msun\leq m_i \leq 99\,\Msun$ and $20\,\Msun \leq M \leq 109\,\Msun$. 

The resulting 90\% confidence upper limits on non-spinning coalescence rates 
are displayed in Figure~\ref{fig:EOBNRUL}, left plot, and in 
Table~\ref{t:UpperLimits}. 
These upper limits supersede those reported in \cite{Collaboration:S5HighMass}.
As explained in the previous section, we used revised priors over rate from the 
S5 analysis. These were obtained using (non-spinning) EOBNRv1 simulated signals,
and a smaller number of IMRPhenomA simulations. 
As described in \cite{Collaboration:S5HighMass} their distances were 
appropriately adjusted to account for the amplitude bias of the older waveform 
families, by comparison with current NR simulations of \ac{BBH} merger.  
Due to the restricted range of parameters of NR simulations available at the 
time of the previous analysis, 
we quote limits on astrophysical rates only for bins within the range of mass
ratio $1\leq q \leq 4$. 
For binaries with both component masses lying between 19 and 28\,$\Msun$ we 
find a 90\% limit of \twentythreetwentythreelimit. 

The averaged sensitive distance to simulated EOBNRv2 waveforms over the 
S6-VSR2/3 observation time, for the entire parameter space of the search, 
is displayed in Figure~\ref{fig:EOBNRUL}, right plot. 

To illustrate our statistical upper limit method we display the cumulative 
posterior probabilities over coalescence rate, for a selection of the mass 
bins of Figure~\ref{fig:EOBNRUL} covering the equal-mass line $m_1=m_2$, 
in Figure~\ref{fig:ratepdfs}. This figure shows the dependence of the quoted
upper limit on the confidence level; we choose to use a 90\% confidence limit
as indicated by the dashed line. 


\subsection{Sensitivity to non-spinning and spinning IMRPhenomB injections}
\label{sec:Phenomrates}
\begin{figure}[tp]
\includegraphics[width=0.95\columnwidth]{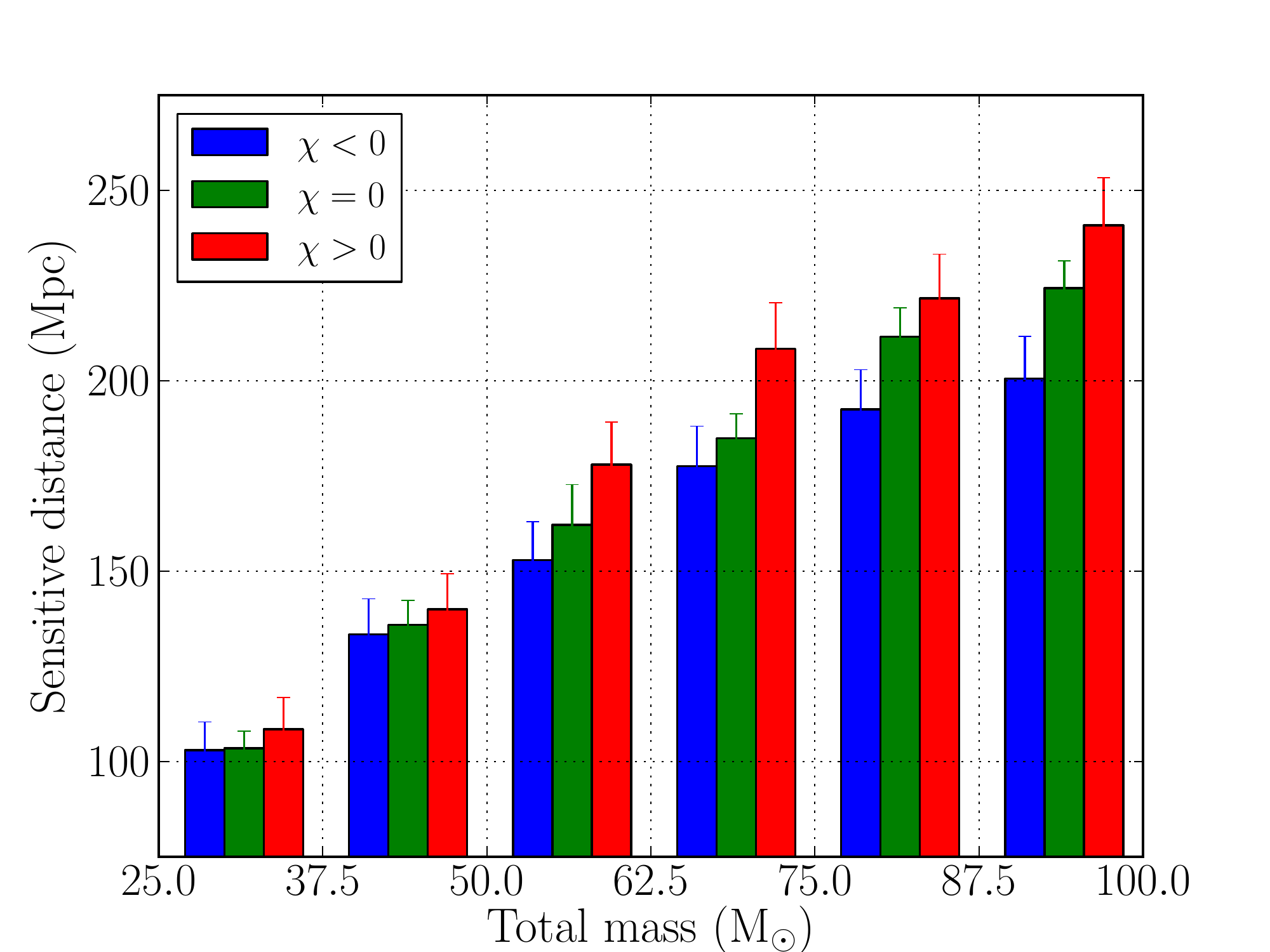}
\caption{Dependence on aligned spin and total mass of the averaged sensitive 
distance of our search to phenomenological inspiral-merger-ringdown 
waveforms. 
For each of 6 bins in total mass $M$, we show the sensitivity for IMRPhenomB 
signals with negative aligned spin parameter $\chi$ (left), non-spinning signals
(centre) and signals with positive aligned spin parameter (right). The simulated 
signal parameters were restricted to mass ratios $1\leq q<4$ and aligned spins 
$-0.85 \leq \chi \leq 0.85$. 
}
\label{fig:PHENOMRANGE}
\end{figure}
The spins of the component black holes are known to have a potentially large
effect on the emitted IMR waveform (e.g.~\cite{Campanelli:2006fy,
Campanelli:2006uy}), affecting the phase evolution and, in the case of spins 
significantly out of alignment with the orbital angular momentum, producing 
amplitude modulations due to precession~\cite{PhysRevD.49.6274}. 
X-ray observations of the spins of accreting black holes in binary systems, 
while technically challenging, indicate a fairly uniform distribution 
over the entire range $0 \le a \equiv S/m^2 \le 1$~\cite{Miller:2009cw, GROJ16:XR, 
McClintock:2006xd, LiuM33:2010, Gou:2009ks, ShaneADT:2006, LiLX:2005}. 
Note that such measurements apply to black holes in X-ray binaries, which may
not be representative of spins in \ac{BBH} systems. 

Indications that spin-orbit misalignment in field binaries may be small come from 
observations of the microquasar XTE J1550-564~\cite{Steiner:2011vr},
and from population synthesis models 
\cite{Fragos:2010tm}. 
For dynamically formed binaries, however, the component spins may be largely 
independent of each other and of the orbital parameters.

In any case it is desirable to perform injections 
using spinning \ac{BBH} coalescence waveforms, in order to estimate how far 
our search was sensitive to such signals. 
Knowledge of spinning \ac{BBH} coalescence waveforms is, however, currently
limited to a relatively small number of numerical relativity simulations (see 
\cite{Ajith:2009bn} and references therein), most having spin 
magnitudes significantly below unity, and only a few including non-aligned 
spins. Only very recently have simulations for near-extreme spins 
\cite{Lovelace:2011nu} been performed. Thus, as a first step towards 
quantifying the sensitivity of the search to spinning waveforms over a broad 
parameter space, we use the IMRPhenomB waveform family \cite{Ajith:2009bn} 
which models IMR signals from \ac{BBH} with aligned/anti-aligned spins. This 
waveform family is parametrized by the total mass $M \equiv m_1 + m_2$, the
mass ratio $q \equiv m_1/m_2$ and a single aligned spin parameter $\chi$,\footnote{This 
parameter is not to be confused with the $\chi^2$ test mentioned earlier in 
Section~\ref{sec:pipeline}.} defined for spins $S_i$ parallel to the orbital 
angular momentum as
\begin{equation}
\label{eq:chi}
 \chi \equiv \frac{m_1}{M} \chi_1 + \frac{m_2}{M} \chi_2,
\end{equation}
where $\chi_i \equiv \mathbf{S}_i \cdot \hat{\mathbf{L}}/m_i^2$ is the 
dimensionless spin of black hole $i$ projected onto the orbital angular 
momentum $\mathbf{L}$. These waveforms are calibrated against 
numerical-relativity simulations in the parameter range $1 \leq q \leq 4$ and 
$-0.85 \leq \chi \leq 0.85$, and, for the inspiral part, to the calculated 
evolution in the extreme-mass-ratio (test mass) limit. 
The waveform family is constructed in the frequency domain and then converted 
to the time domain via an inverse discrete Fourier transform. 

As simulated signals we used two sets of IMRPhenomB injections, a non-spinning 
set and a spinning set. Both were uniformly distributed in total mass between 
25 and 100\,\Msun, and uniformly distributed in $q/(q+1) \equiv m_1/M$ for a 
given $M$, between the limits $1\leq q<4$. 
In addition, the spinning 
injections were assigned aligned spin components $\chi_i$ uniformly distributed 
between -0.85 and 0.85.

To illustrate the effect of aligned spin on the search sensitivity, we plot in
Figure~\ref{fig:PHENOMRANGE} the average sensitive distance over the S6-VSR2/3 
observation time, in bins of total mass $M$, for both non-spinning simulated 
signals and for injections with $\chi<0$ and $\chi>0$ respectively. 

Component spin is expected to have several effects on our search, compared to
its performance for non-spinning systems. First, the amplitude of the expected 
signal from a coalescence at a given distance may depend on spin: see for instance
Figure~3 of \cite{Ajith:2009bn}, where the horizon distance for IMRPhenomB signals 
with optimally fitting templates, with Initial LIGO noise spectra, 
was found to increase steeply 
with increasing positive $\chi$. Second, the EOBNR templates used in our search
may have reduced overlap with the simulated spinning signals, leading to a loss
of sensitivity. Third, the signal-based $\chi^2$ test values are expected to be 
higher than if exactly matched spinning templates were used, due to ``un-matched'' 
excess power in the signals; this would further reduce the search sensitivity. 
Given the complexity of the search
pipeline, it is not clear which effect would dominate. Figure~\ref{fig:PHENOMRANGE}
indicates higher sensitivity to positive-$\chi$ signals even with the current 
non-spinning templates, but also shows that the search is significantly less 
sensitive to negative-$\chi$ signals at higher values of total mass $M$. 


\subsection{Waveforms including higher spherical harmonic modes}
In the filter templates and in all injections used in this search, we consider 
only the dominant mode of GW emission from coalescing binaries, the $(l,m) = (2,2)$ 
mode. In general, higher-order modes are important in \ac{BBH} with asymmetric 
masses and significant component spins. Omitting these modes in templates will 
neglect their contributions to the SNR \cite{VanDenBroeck:2006qu,McKechan:2011ps}, 
and may also lead to a worse (higher) value of the $\chi^2$ test, tending to 
reduce the sensitivity of the search.  
However, such effects will depend strongly on the mass ratio and on extrinsic 
(angular) parameters, and it is beyond the scope of this analysis to 
investigate them in detail. 

In~\cite{Pan:2011gk} the mismatch between NR waveforms containing the strongest 
$7$ modes observed with a binary inclination angle of $\pi/3$, and EOBNRv2 
templates containing only the $(l, m)=(2,2)$ mode, was calculated using the 
Advanced LIGO zero-detuning high-power PSD \cite{PSD:AL}. This mismatch 
took values up to 10\% for BBHs with mass ratio $q=1$--$6$ and total mass 
$M<100\Msun$. Adding non-dominant modes to the EOBNRv2 waveforms reduced the 
mismatch to below 1\% (for the same inclination angle and a template containing 5 
modes).

\subsection{Astrophysical implications}

There is no commonly accepted astrophysical rate estimate for high-mass 
\acp{CBC}, owing to the many possible formation scenarios and the 
considerable uncertainties affecting them. In \cite{Bulik:2008,Belczynski:2011qp} 
a rate of $3.6^{+5.0}_{-2.6}\times 10^{-7}\,\perMpcyr$
for IMR signals from binaries with chirp mass comparable to $15\,\Msun$ 
was estimated based on two observed tight binaries believed to consist of
a massive stellar BH and a Wolf-Rayet star. Our 90\% upper rate limit for
the bin with component masses $19<m_1/\Msun<28$, $10<m_2/\Msun<19$, for 
which the chirp mass ranges between $12$ and $20\,\Msun$, is 
\twentythreefourteenlimit. Thus, current searches are close to the
sensitivity necessary to put nontrivial constraints on astrophysical 
scenarios of \ac{BBH} formation and evolution. However, we remind the 
reader that systems with near-extremal ($S_i/m_i^2>\!0.85$) or 
significantly non-aligned spins, or for which higher signal harmonics 
make a considerable contribution to the waveform seen at the detectors, 
were not included in our sensitivity studies.

\section{Discussion}
\label{sec:discussion}
\noindent
The present search is an advance over that reported in 
\cite{Collaboration:S5HighMass} in three main respects: the improved
sensitivity of the LIGO and Virgo detectors over previous science runs; 
improved understanding of the search background 
by identifying the duration of the IMR templates as the 
dominant parameter controlling their output in non-Gaussian detector data; 
and the use of updated, more accurate signal models to assess search 
sensitivity, including models describing component spins aligned to the 
orbital angular momentum. 

There are, however, many issues that remain to be addressed in order for 
future data from advanced detectors \cite{0264-9381-27-8-084006,Accadia:2011} 
to be best exploited in searching for high-mass \ac{CBC}. Among these, the 
metric currently used for template bank placement and for testing mass
coincidence between detectors is based on the inspiral portion of \ac{CBC} 
waveforms only. The search may be improved by using a more accurate metric 
for IMR waveforms, and more radically by also including spin-aligned IMR 
templates as matched filters, which may significantly increase the 
sensitivity to spinning \ac{BBH}. 

Separating signal from non-Gaussian noise events in short filter templates, 
where our signal-based $\chi^2$ tests are not effective, remains a difficult
problem. Improved methods, including the use of amplitude consistency tests 
between different detectors and multivariate classifiers, are currently 
being investigated.

Advanced detectors coming on line in the coming years will
improve the sensitivity to GW 
relative to the first generation by a factor 10 over a broad frequency 
range, and will achieve good sensitivity down to a low-frequency limit of 
$\sim\! 10$\,Hz. The volume of space over which future searches will be 
sensitive to IMR signals is therefore expected to increase by a factor 
$10^3$ or more depending on the binary masses. We thus expect to extract 
significant information on \ac{BBH} source populations over the parameter
space of future searches.

\acknowledgments
The authors gratefully acknowledge the support of the United States
National Science Foundation for the construction and operation of the
LIGO Laboratory, the Science and Technology Facilities Council of the
United Kingdom, the Max-Planck-Society, and the State of
Niedersachsen/Germany for support of the construction and operation of
the GEO600 detector, and the Italian Istituto Nazionale di Fisica
Nucleare and the French Centre National de la Recherche Scientifique
for the construction and operation of the Virgo detector. The authors
also gratefully acknowledge the support of the research by these
agencies and by the Australian Research Council, 
the International Science Linkages program of the Commonwealth of Australia,
the Council of Scientific and Industrial Research of India, 
the Istituto Nazionale di Fisica Nucleare of Italy, 
the Spanish Ministerio de Econom\'ia y Competitividad,
the Conselleria d'Economia Hisenda i Innovaci\'o of the
Govern de les Illes Balears, the Foundation for Fundamental Research
on Matter supported by the Netherlands Organisation for Scientific Research, 
the Polish Ministry of Science and Higher Education, the FOCUS
Programme of Foundation for Polish Science,
the Royal Society, the Scottish Funding Council, the
Scottish Universities Physics Alliance, The National Aeronautics and
Space Administration, 
the National Research Foundation of Korea,
Industry Canada and the Province of Ontario through the Ministry of Economic Development and Innovation, 
the National Science and Engineering Research Council Canada,
the Carnegie Trust, the Leverhulme Trust, the
David and Lucile Packard Foundation, the Research Corporation, and
the Alfred P. Sloan Foundation.


\end{document}